\documentclass[aps,prx,twocolumn,superscriptaddress,nofootinbib,longbibliography]{revtex4-2}
\usepackage[utf8]{inputenc}
\usepackage[colorlinks=true,citecolor=blue,linkcolor=blue,urlcolor=blue]{hyperref}
\usepackage{amssymb}
\usepackage[version=4]{mhchem}
\usepackage{braket}
\usepackage{bm}
\usepackage[svgnames]{xcolor}
\usepackage{natbib}
\usepackage{graphicx, nicefrac, textcomp}
\usepackage{soul}
\usepackage{amsmath}
\usepackage{booktabs}
\usepackage{comment}

\usepackage{algorithm}
\usepackage{algpseudocode}
\usepackage{float}
\usepackage{enumitem}

\begin{document}
\title{Natural-Orbital-Based Neural Network Configuration Interaction}

\author{Louis Thirion}
\email{louis.thirion@fau.de}
\affiliation{Department of Physics, Friedrich-Alexander-Universit\"at Erlangen-N\"urnberg, 91058 Erlangen, Germany}
\affiliation{Science Institute and Faculty of Physical Sciences, University of Iceland, 107 Reykjav\'ik, Iceland}
\author{Yorick L.\ A.\ Schmerwitz}
\affiliation{Max-Planck-Institut f\"ur Kohlenforschung, 45470 M\"ulheim an der Ruhr, Germany}
\author{Max Kroesbergen}
\affiliation{Department of Physics, Friedrich-Alexander-Universit\"at Erlangen-N\"urnberg, 91058 Erlangen, Germany}
\author{Gianluca Levi}
\affiliation{Department of Chemical and Pharmaceutical Sciences, University of Trieste, 34127 Trieste, Italy}
\affiliation{Science Institute and Faculty of Physical Sciences, University of Iceland, 107 Reykjav\'ik, Iceland}
\author{Elvar~\"O.~J\'onsson}
\affiliation{Science Institute and Faculty of Physical Sciences, University of Iceland, 107 Reykjav\'ik, Iceland}
\author{Pavlo Bilous}
\affiliation{Max Planck Institute for the Science of Light, 91058 Erlangen, Germany}
\author{Hannes J\'onsson}
\email{hj@hi.is}
\affiliation{Science Institute and Faculty of Physical Sciences, University of Iceland, 107 Reykjav\'ik, Iceland}
\author{Philipp Hansmann}
\email{philipp.hansmann@fau.de}
\affiliation{Department of Physics, Friedrich-Alexander-Universit\"at Erlangen-N\"urnberg, 91058 Erlangen, Germany}
\affiliation{Science Institute and Faculty of Physical Sciences, University of Iceland, 107 Reykjav\'ik, Iceland}

\date{\today}

\begin{abstract}
Selective configuration interaction methods approximate correlated molecular ground- and excited states by considering only the most relevant Slater determinants in the expansion. 
While a recently proposed neural-network-assisted approach efficiently identifies such determinants, the procedure typically relies on canonical Hartree-Fock orbitals, which are optimized only at the mean-field level. 
Here we assess approximate natural orbitals -- eigenfunctions of the one-particle density matrix computed from intermediate many-body eigenstates -- as an alternative.
Across our benchmarks for \ce{H2O}, \ce{NH3}, \ce{CO}, and \ce{C3H8} we see a consistent reduction in the required determinants for a given accuracy of the computed correlation energy compared to full configuration interaction calculations. 
Our results confirm that even approximate natural orbitals constitute a simple yet powerful strategy to enhance the efficiency of neural-network-assisted configuration interaction calculations.
\end{abstract}

\maketitle

\section{Introduction}
Accurate computation of molecular ground-state energies is challenging because the formally exact full configuration interaction (CI) solution becomes intractable as the configuration space grows \cite{Szabo_ModnQChem_2012}. 
Hartree-Fock offers a compact mean-field reference; the missing electron correlation is commonly partitioned into dynamic, often well captured by M{\o}ller--Plesset perturbation theory (MP$n$) and coupled cluster (CC) methods \cite{Moller1934,Bartlett2007}, and static, which calls for multireference treatments such as complete active space self-consistent field (CASSCF) and complete active space second-order perturbation theory (CASPT2) \cite{Roos1980,Andersson1992}.
CI represents the wave function in a basis of Slater determinants. 
Selective CI (SCI) methods retain only the most important determinants, yielding compact and systematically improvable expansions. 
Classical and modern realizations - configuration interaction using a perturbative selection made iteratively (CIPSI), heat-bath configuration interaction (HCI), semistochastic HCI, and adaptive sampling configuration interaction (ASCI) - differ in selection criteria and the treatment of the residual space, but all deliver high accuracy at reduced cost \cite{CIPSI,HCI,SHCI1,SHCI2,ASCI}.
Machine learning (ML) has recently introduced a new twist and has been used to accelerate determinant selection by predicting importance directly. 
Neural classifiers trained during the CI iterations, as in MLCI and active-learning CI, reduce the number of explicitly evaluated determinants while approaching full CI accuracy \cite{Coe_MLCI_JChemTC_2018,Jeong_ALCI_JChemTC_2021}.\\

Building on this idea, we recently introduced a neural-network‑assisted SCI (NNCI) framework that trains a classifier on-the-fly to identify relevant determinants during SCI iterations in condensed matter and quantum chemistry contexts \cite{Bilous2025,N2_JCTC, SOLAX}. 
NNCI turns out to accurately recover correlation energies with a small fraction of determinants, and our prior results indicate that operating in larger single‑particle bases can be much more beneficial than trying to reach the full CI solution in a small basis \cite{N2_JCTC}. 
This observation motivates coupling SCI with orbital optimization on the single-particle level.
A practical limitation of many SCI approaches, including NNCI, is the use of canonical Hartree-Fock molecular orbitals (MOs), which are optimized only at the mean‑field level and do not necessarily yield the most compact many‑body expansions. 
Natural orbitals (NOs) - eigenfunctions of the one‑particle reduced density matrix (1-RDM) of a correlated state- typically provide a more natural starting point for correlated calculations and accelerate CI convergence. 
Importantly, they can be approximated inexpensively from preliminary correlated calculations.\\

The present work quantifies the impact of using NOs as the single‑particle basis for a many-body NNCI treatment. 
We generate approximate NOs from intermediate NNCI solutions and continue the selection in the rotated basis. 
Benchmarks for \ce{H2O}, \ce{NH3}, \ce{CO}, and \ce{C3H8} demonstrate that, for a fixed active space, NOs consistently and significantly reduce the number of determinants required to reach a target accuracy compared with Hartree-Fock orbitals. 
We provide practical guidance for constructing these orbitals and for integrating single‑particle basis optimization into ML‑accelerated SCI workflows.

\section{Methods}

\subsection{Electronic-structure setup}
The single-particle basis for the configuration interaction Hamiltonian is generated with GPAW~\cite{GPAW2024}. 
Hartree-Fock calculations are carried out in the plane-wave mode with a 1000\,eV cutoff and a uniform real-space grid of 0.18\,\AA. 
Convergence is reached when the squared residual of the Hartree-Fock equations falls below $10^{-11}\,\text{eV}^{2}$ per valence electron. 
GPAW employs the projector-augmented wave (PAW) formalism \cite{paw1,paw2} together with the frozen-core approximation, i.e.\ scalar relativistic atomic calculations provide the core electrons, which are not explicitly considered in the subsequent Hartree-Fock and NNCI steps.
These settings yield the canonical Hartree-Fock occupied orbitals, while leaving the virtual space undefined. 
In principle, as many orbitals exist as plane waves in the basis. 
In practice, only $\sim\!100$ high-quality orbitals are tractable for many-body calculations. 
Occupied orbitals are initialized from predefined atomic basis functions, evaluated on the real-space grid, Fourier-transformed, orthonormalized, and then optimized by direct minimization \cite{dm1,dm2}. 
Virtual orbitals are generated analogously from the remaining atomic basis functions, but without optimization. 
The periodic cell is chosen sufficiently large to accommodate all basis functions up to their cutoff values in real space.

For propane, numerical atomic orbital (NAO) basis sets \cite{Larsen2009} are used: four $s$ and two $p$ functions for hydrogen; four $s$, three $p$, and two $d$ for carbon. 
For ammonia, water and carbon monoxide, mixed NAO/Gaussian bases are employed: one NAO $s$ for hydrogen, and one NAO $s$ plus one $p$ for carbon, nitrogen, and oxygen.
Gaussian primitives are taken from the aug-cc-pVTZ basis, comprising three $s$, three $p$, and two $d$ functions for hydrogen, and three $s$, three $p$, three $d$, and two $f$ functions for carbon, nitrogen, and oxygen.

This results in a set of single-particle orbitals $\{\varphi_i(\bm r)\}$ for $i\leq \lceil N_\mathrm{el}/2\rceil$ being the canonical Hartree-Fock orbitals and else the virtual orbitals.
To construct the Hamiltonian for a subsequent many-body calculation, we compute the single- and two-particle integrals
\begin{align}
t_{ij} &\equiv \int \text{d}\mathbf{r} \varphi^*_i(\mathbf{r}) \left(-\frac{1}{2}\nabla^2 + V^\text{eff}(\mathbf{r}) \right) \varphi_j(\mathbf{r})\\
U_{ijkl} &\equiv \int \text{d}\mathbf{r}\,\text{d}\mathbf{r'} \varphi^*_i(\mathbf{r}) \varphi_j(\mathbf{r}) \frac{1}{|\mathbf{r}-\mathbf{r}'|}  \varphi_k^*(\mathbf{r}') \varphi_l(\mathbf{r}')
\end{align}
where $V_\text{eff}(\mathbf{r})$ is the self-consistent mean-field potential.

The electronic Hamiltonian for the many-body system is composed as follows:
\begin{equation}
H = H_0 + H_\text{int} - \text{MF}[H_\text{int}],
\label{eq:fullH}
\end{equation}
with
\begin{align}
H_0 &= \;\sum_{i,j,\sigma} \;\; t_{ij} \;\;\;\; c_{i\sigma}^\dagger c^{\phantom{\dagger}}_{j\sigma}\,, \\
H_\text{int} &= \,\sum_{\substack{i,j,k,l \\ \sigma,\sigma'}} \,U_{ijkl} \;c_{i\sigma}^\dagger c_{j\sigma}^{\phantom{\dagger}} c^\dagger_{k\sigma'} c^{\phantom{\dagger}}_{l\sigma'}\,,
\end{align}
using the one- and two-particle integrals and $c_{i\sigma}^\dagger$ ($c_{i\sigma}$) being the creation (annihilation) operator for spin $\sigma$ in molecular orbital $i$. 
$\mathrm{MF}[H_\text{int}]$ denotes the mean-field decoupled interaction used to remove double counting.

\subsection{NNCI algorithm} 
Full details are provided in Refs.~\cite{Bilous2025, N2_JCTC}. Here we only provide a compact summary. 
The algorithm combines \emph{active learning} with \emph{iterative determinant selection} to construct an efficient CI basis for correlated electron systems. 
Starting from a small initial determinant set, typically the Hartree-Fock solution, the Hamiltonian is repeatedly applied to generate new candidate configurations, which we denote as the pool.
A convolutional \emph{neural network classifier} that is trained on information obtained from a smaller exact diagonalization on determinants sampled from the pool predicts the importance of new determinants based on their occupation-number representation. 
Determinants identified as \emph{important} are added to the CI basis, and the process is iterated until convergence of the target observable, typically the ground-state energy. 
This adaptive cycle systematically refines the determinant space, significantly reducing the computational cost compared to non-ML truncation schemes while maintaining high accuracy \cite{Bilous2025}. The NNCI framework is general and can be applied to diverse quantum-clusters or molecular Hamiltonians.

Application of this algorithm finds a compact representation of the exact many-body ground state $|\Psi_\text{gs}\rangle$ that is expanded in Slater determinants $|\phi\rangle$ as
\begin{equation}
|\Psi^\mathrm{ex}_\text{gs}\rangle = \sum_{\phi \in \mathcal{H}} c_\phi |\phi\rangle \approx \sum_{\phi \in \mathcal{H}^\mathrm{s}} \tilde{c}_\phi |\phi\rangle,
\end{equation}
with $\mathrm{dim}\left(\mathcal{H}^\mathrm{s}\right) \ll \mathrm{dim}\left(\mathcal{H}\right)$ for the selected subspace \mbox{$\mathcal{H}^\text{s}$} of the full Hilbert space $\mathcal{H}$.

The total energy $E$ is the expectation value of the full Hamiltonian~\eqref{eq:fullH}. 
We define the correlation energy as
\begin{equation}
  E_\mathrm{corr} \equiv E - E_\mathrm{HF},
  \label{eq:Ecorr-def}
\end{equation}
where $E_\mathrm{HF}$ is the Hartree-Fock energy evaluated for the same one-electron basis.

\subsection{Optimal bases for SCI}
We define a single-particle basis to be \emph{optimal} for SCI calculations when it renders the variational many-body expansion as compact as possible for the targeted state — i.e., it minimizes the number of Slater determinants required to achieve a chosen accuracy. 
Consequently, an optimal basis accelerates convergence of the SCI energy with respect to the number of retained determinants. 
Additionally, important determinants should be \emph{connected} by low orders of the Hamiltonian so that the iterative series converges rapidly and the dominant configurations are found early. 
\emph{Natural orbitals} (NOs), defined as eigenfunctions of the 1-RDM, are among the most widely used optimal bases because they concentrate occupation (and correlation effects) and often yield the fastest CI convergence - this idea, introduced by L\"owdin \cite{Lowdin1955}, remains central to many modern many-body methods.

Because the precise definition of an optimal basis depends on the metric (energy error, determinant count, locality, entanglement), a variety of orbital-optimization strategies exists. 
State-specific optimization within multiconfigurational self-consistent-field procedures (MCSCF) tailors orbitals to the static-correlation structure of the active space, while direct orbital rotations that minimize the SCI energy (orbital-optimized SCI) can further reduce the variational space required for chemical accuracy \cite{Yao2021}. 

In practice, NOs are an excellent default when compactness of the variational wave function is the primary goal, whereas localized or pair-optimized orbitals may be advantageous when spatial locality dominates. 
A pragmatic strategy — especially in iterative or ML-assisted SCI workflows such as NNCI — is to obtain a preliminary correlated state, compute its NOs, and resume selection in the rotated basis, which has repeatedly been found to produce more compact expansions \cite{Yao2021}.

\subsection{Natural orbitals}
NOs are defined as eigenfunctions of the one-particle reduced density matrix (1-RDM) of the target many-body state. 
\begin{equation}
  \rho^{(1)}_{ij}=\bra{\Psi_{\mathrm{gs}}}\,c_i^\dagger c_j\,\ket{\Psi_{\mathrm{gs}}}\,,
  \label{eq:opdm-def}
\end{equation}
where $i$ and $j$ index the orbital basis.
By construction, NOs redistribute correlation into fractional occupations, concentrating weight into a smaller set of orbitals and thereby often producing substantially more compact CI expansions than canonical Hartree-Fock orbitals. 
This compactness reduces the number of determinants needed to meet a given accuracy.\\

However, calculation of the exact NOs requires the fully correlated 1-RDM and are therefore unavailable \emph{a priori}. 
In practice one employs \emph{approximate} NOs obtained from a lower-effort many-body calculation. 
Such approximate NOs typically retain the favorable convergence properties of exact NOs while remaining inexpensive to construct, and are widely used as a pragmatic compromise \cite{Yao2021}.\\

\begin{figure*}[htb]
    \includegraphics[width=\textwidth]{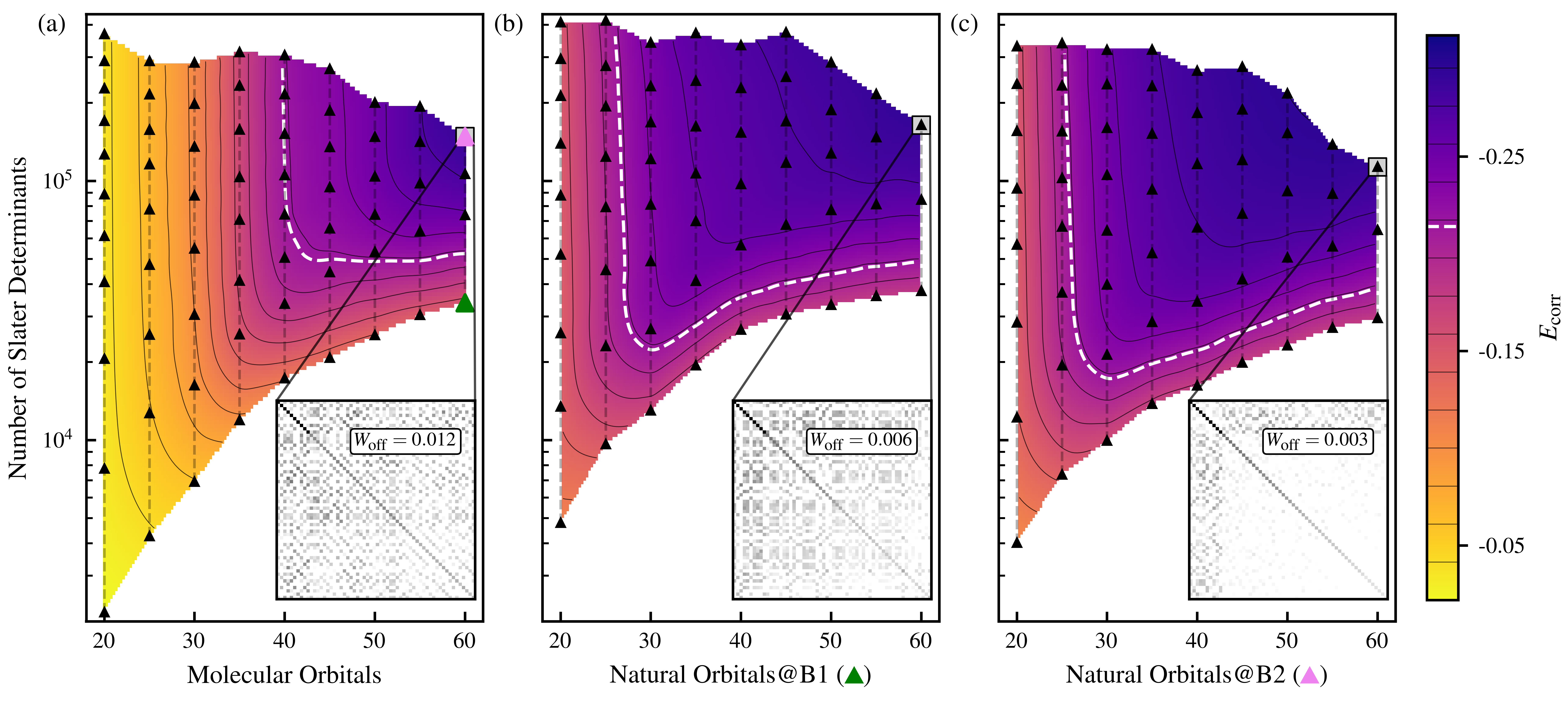} 
    \caption{Color map for the correlation energy $E_\mathrm{corr}$ of \ce{C3H8} as a function of the number of determinants (vertical axis) and the number of basis orbitals (horizontal axis) with thin black iso-contour lines. (a) molecular orbital basis and (b) natural orbitals obtained from an intermediate many-body solution (green symbol in (a)) and (c) natural orbitals obtained from a costly many-body solution (gray symbol in (a)). The white dashed line indicates full-CI benchmarks from Ref.~\cite{Gao2024}. The inset displays the absolute values of the 1-RDM at the best converged point in each basis, its off-diagonal weight  $W_\mathrm{off}$ (see Eq.~\eqref{eq:off_diag_density_norm}) indicates the quality of the basis as $W_\mathrm{off}\rightarrow 0$ for the exact solution.}
    \label{fig:E_corr_C3H8_molecular_orbitals}
\end{figure*}
\begin{figure*}[!htb]
    \includegraphics[width=\textwidth]{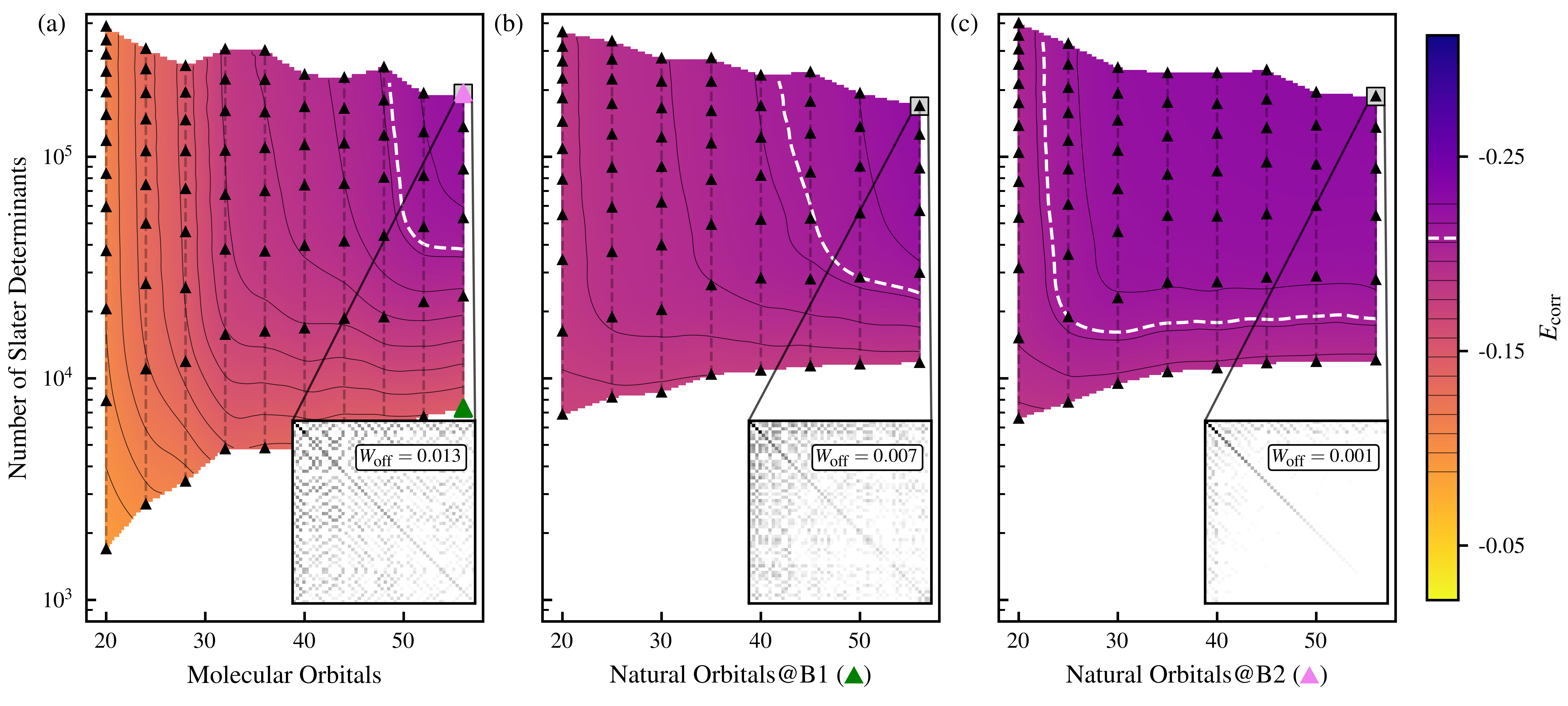} 
    \caption{Color map for the correlation energy $E_\mathrm{corr}$ of \ce{NH3} as a function of the number of determinants (vertical axis) and the number of basis orbitals (horizontal axis) with thin black iso-contour lines. (a) molecular orbital basis and (b) natural orbitals obtained from an intermediate many-body solution (green symbol in (a)) and (c) natural orbitals obtained from a costly many-body solution (gray symbol in (a)). The white dashed line indicates full-CI benchmarks from Ref.~\cite{Gao2024}. The inset displays the absolute values of the 1-RDM at the best converged point in each basis, its off-diagonal weight  $W_\mathrm{off}$ (see Eq.~\eqref{eq:off_diag_density_norm}) indicates the quality of the basis as $W_\mathrm{off}\rightarrow 0$ for the exact solution.}
    \label{fig:E_corr_NH3_molecular_orbitals}
\end{figure*}

\begin{figure*}[htb]
    \includegraphics[width=\textwidth]{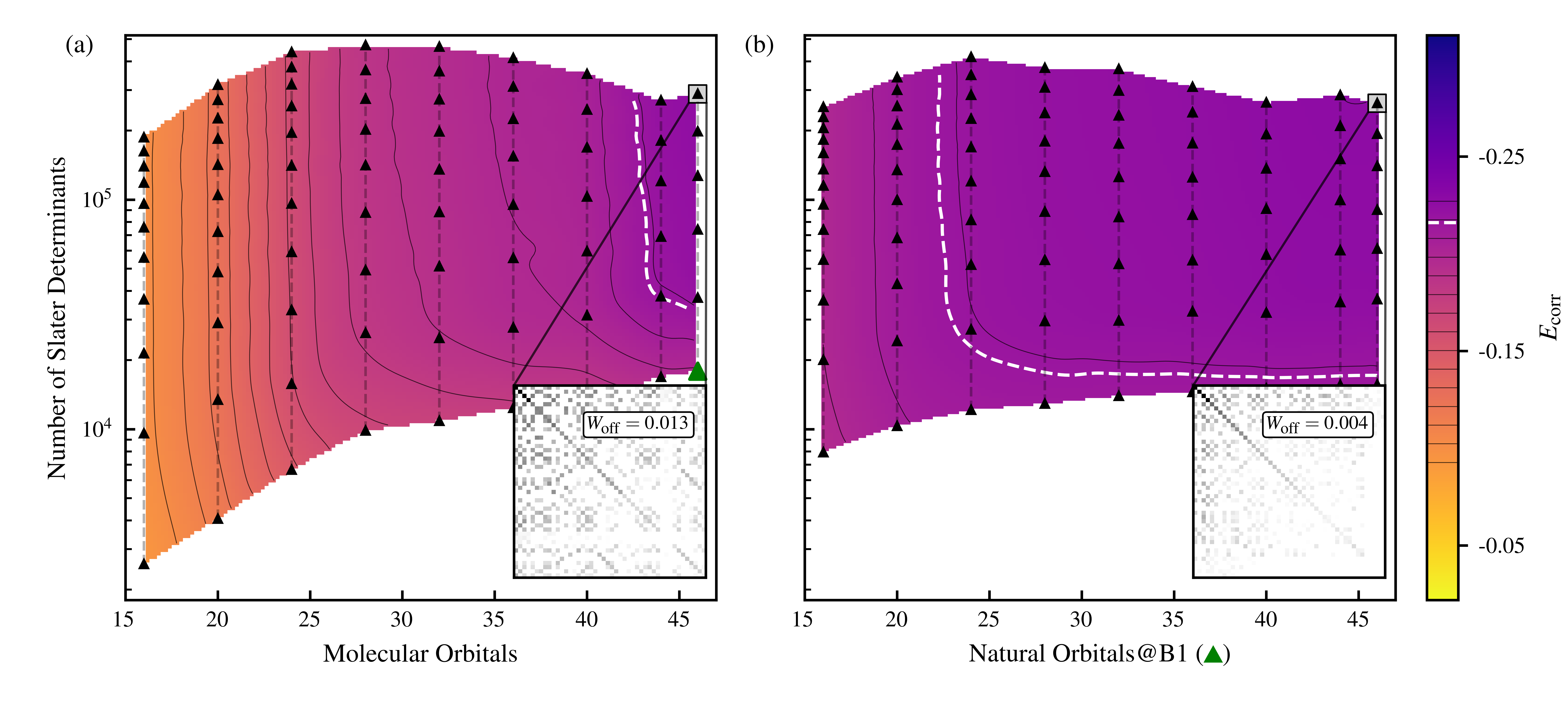} 
    \caption{Color map for the correlation energy $E_\mathrm{corr}$ of \ce{H2O} as a function of the number of determinants (vertical axis) and the number of basis orbitals (horizontal axis) with thin black iso-contour lines. (a) molecular orbital basis and (b) natural orbitals obtained from an intermediate many-body solution (green symbol in (a)). The white dashed line indicates full-CI benchmarks from Ref.~\cite{Gao2024}. The inset shows the absolute values of the 1-RDM at the best converged point in each basis, its off-diagonal weight  $W_\mathrm{off}$ (see Eq.~\eqref{eq:off_diag_density_norm}) indicates the quality of the basis as $W_\mathrm{off}\rightarrow 0$ for the exact solution.}
    \label{fig:E_corr_H2O_molecular_orbitals}
\end{figure*}
\begin{figure*}[!htb]
    \includegraphics[width=\textwidth]{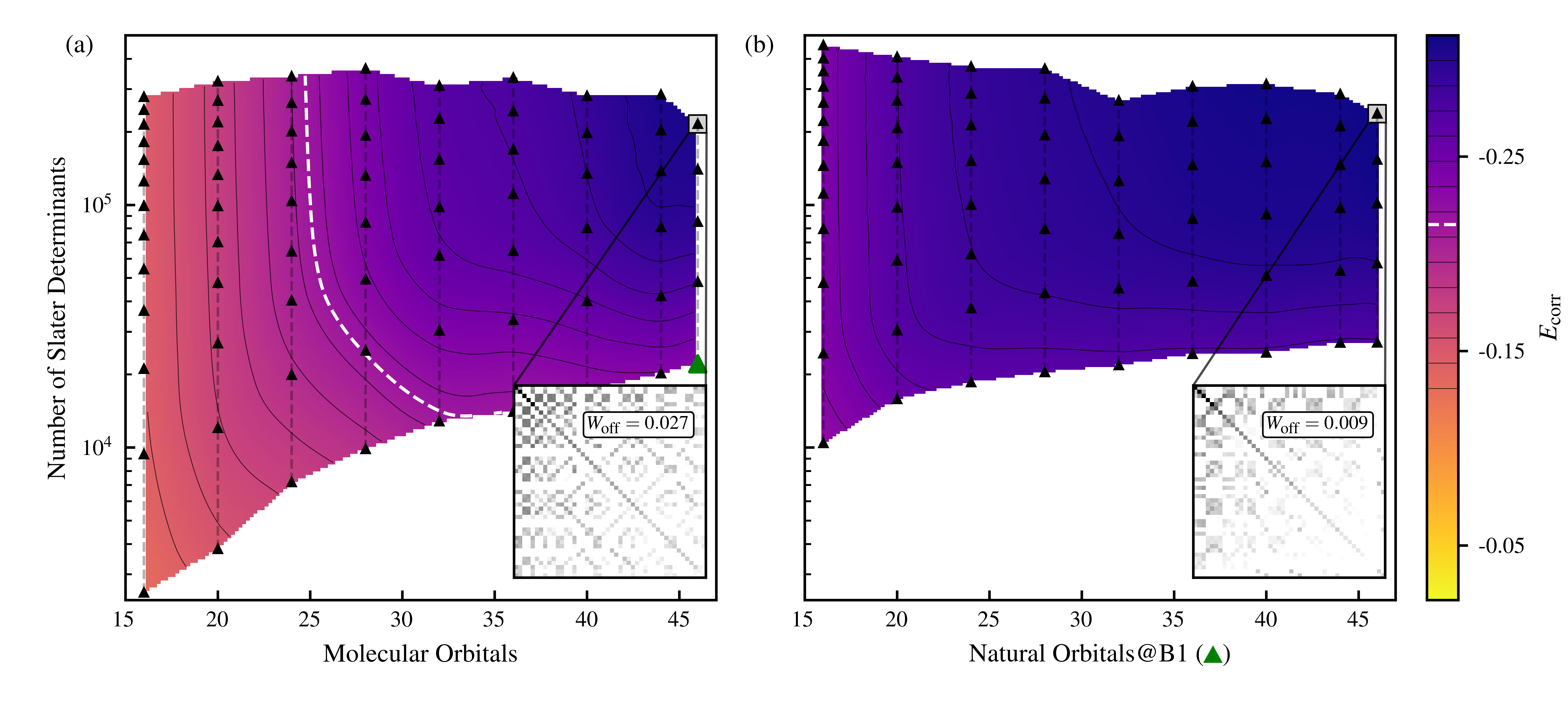} 
    \caption{Color map for the correlation energy $E_\mathrm{corr}$ of CO as a function of the number of determinants (vertical axis) and the number of basis orbitals (horizontal axis) with thin black iso-contour lines. (a) molecular orbital basis and (b) natural orbitals obtained from an intermediate many-body solution (green symbol in (a)). The white dashed line indicates full-CI benchmarks from Ref.~\cite{Gao2024}. The inset shows the absolute values of the 1-RDM at the best converged point in each basis, its off-diagonal weight  $W_\mathrm{off}$ (see Eq.~\eqref{eq:off_diag_density_norm}) indicates the quality of the basis as $W_\mathrm{off}\rightarrow 0$ for the exact solution.}
    \label{fig:E_corr_CO_molecular_orbitals}
\end{figure*}

\paragraph{Definition and notation -}
For a correlated many-body ground state $\ket{\Psi_{\mathrm{gs}}}$, the 1-RDM is given by~\eqref{eq:opdm-def}. 
In spin-restricted cases we use the spin-summed 1-RDM. The NOs $\{\tilde\psi_p\}$ follow from the unitary transformation $R$ that makes $\rho^{(1)}$ diagonal:
\begin{align}
  \tilde\psi_p(\mathbf r) &= \sum_i R_{ip}\,\psi_i(\mathbf r)\,,
  \label{eq:no-rotation}
  \\
  \tilde\rho^{(1)}_{pq} &= \sum_{ij} R_{ip}^*\,\rho^{(1)}_{ij}\,R_{jq}
  \;=\; n_p\,\delta_{pq}\,,
\end{align}
with natural occupations $0\le n_p\le 1$ in a spin-orbital basis (or $0\le n_p\le 2$ for the spin-summed, restricted case). 
The creation and annihilation operators transform as
\begin{align}
  \tilde c_p^\dagger &= \sum_i R_{ip}^*\, c_i^\dagger\,, \label{eq:op-transform-dag} \\
  \tilde c_p          &= \sum_i R_{ip}\, c_i\,.       \label{eq:op-transform}
\end{align}
Hence, the one- and two-electron integrals transform as

\begin{align}
  \tilde h_{pq}   &= \sum_{ij} R_{ip}^*\, R_{jq}\, h_{ij}\, ,\\
  \tilde U_{pqrs} &= \sum_{ijkl} R^*_{ip}\, R_{jq}\, R^*_{kr}\, R_{ls}\, U_{ijkl}\,.
  \label{eq:integral-rotation}
\end{align}
If $\ket{\Psi_{\mathrm{gs}}}$ is replaced by the Hartree-Fock ground state, the 1-RDM is diagonal in the MO basis and the NOs coincide with the canonical Hartree-Fock orbitals.

The choice of the single-particle basis affects how compactly the correlated wave function is represented.
A 1-RDM $\rho^{(1)}$ that is nearly diagonal in a given basis indicates that most of the correlation is captured by the diagonal occupations, and the CI expansion requires fewer off-diagonal excitations to represent the state accurately. 
To quantify this notion, we define the \emph{normalized off-diagonal weight} of a 1-RDM $\rho^{(1)}$ as
\begin{equation}\label{eq:off_diag_density_norm}
W_\mathrm{off} = \frac{\left\|\rho_\mathrm{off}^{(1)}\right\|^2}{\left\|\rho^{(1)}\right\|^2} = \frac{\sum_{i\neq j} \left|\rho_{ij}^{(1)}\right|^2}{\sum_{ij} \left|\rho_{ij}^{(1)}\right|^2}
\end{equation}
This measure takes values between $0$ and $1$, where $0$ corresponds to the optimal, fully diagonal 1-RDM (most compact representation in the chosen basis) and values $W_\mathrm{off}>0$ signal a less compact CI expansion.\\

\paragraph{Practical implementation -}
\emph{Exact} NOs would require the exact ground state in the chosen one-electron basis (i.e., the full CI limit within the active space) and are therefore known only \emph{a posteriori}. 
We instead construct \emph{approximate} NOs from NNCI ground-state approximations at specified convergence levels. 
We denote by $\mathrm{NO@}B$ the natural orbitals obtained by: (i) running NNCI with a determinant budget $B$, (ii) forming $\rho^{(1)}$ from that state, (iii) diagonalizing once, and (iv) performing a single integral rotation, 
Eqs.~\eqref{eq:no-rotation}–\eqref{eq:integral-rotation}, before continuing selection in the rotated basis. 
The overhead beyond the NNCI solve is a diagonalization of an $N_\mathrm{orb}\!\times\!N_\mathrm{orb}$ matrix and one four-index integral rotation. 
In the presented figures and tables we compare Hartree-Fock orbitals to $\mathrm{NO@}B_1$ (moderate budget) and $\mathrm{NO@}B_2$ (higher budget), which approach the exact-NO limit as the underlying NNCI state improves.

\section{Results}
Our results are summarized in Table~\ref{tab:Ecorr} and in Figures \ref{fig:E_corr_C3H8_molecular_orbitals}, \ref{fig:E_corr_NH3_molecular_orbitals}, \ref{fig:E_corr_H2O_molecular_orbitals}, and \ref{fig:E_corr_CO_molecular_orbitals}. 
Unless stated otherwise, the color maps encode the correlation energy $E_\mathrm{corr}$ as defined in Eq.~\eqref{eq:Ecorr-def} (more negative $\Rightarrow$ darker). 
Within each figure, all panels use an identical colorbar range to enable a direct visual comparison between bases. 
The white dashed lines in the plots indicate full CI benchmarks taken from Ref.~\cite{Gao2024}.
The insets display the magnitude of the 1-RDM in the working orbital basis (linear scale) together with the off-diagonal weight $W_\mathrm{off}$ defined in \eqref{eq:off_diag_density_norm}.
Lower $W_\mathrm{off}$ indicates a more diagonal 1-RDM and hence also a single-particle basis that is closer to the real NOs. This also results in a more compact representation of the many-body wave function expansion.\\

\paragraph{\ce{C3H8} -}
Fig.~\ref{fig:E_corr_C3H8_molecular_orbitals} shows a pronounced efficiency gain when moving from Hartree-Fock orbitals to NO@B$_1$ and further to NO@B$_2$.
More specifically, we observe improvement along both axes: at fixed $N_{\rm orb}$, fewer determinants are required, and at fixed $N_{\rm det}$, smaller orbital spaces already become effective.
Quantitatively, the best achievable energy across the scan improves from $-0.2895$\,Ha (Hartree-Fock) to $-0.2913$\,Ha (NO@B$_1$) to $-0.2954$\,Ha (NO@B$_2$), i.e., a net gain of $\sim\!6$\,mHa over Hartree-Fock (Table~I).
Moreover, for \ce{C3H8} we observe a `sweet-spot' ridge (positive slope of the contour lines above $\approx30$ orbitals) which reflects that, at fixed target $E_\mathrm{corr}$, there exists an optimal orbital space that minimizes the determinant budget. 
For the other systems we do not observe such behavior and get flat or even negaitive slope contours.  

The insets show the increasing diagonal dominance quantified by $W_\mathrm{off}$ and consistent with recovering more correlation with a comparably compact determinant expansion built on natural orbitals.\\

\paragraph{\ce{NH3} -}
For \ce{NH3} (Fig.~\ref{fig:E_corr_NH3_molecular_orbitals}), NO@B$_1$ we find the same result and efficiency gains for the computation of the correlation energy (best energies: $-0.2217$\,Ha $\rightarrow$ $-0.2243$\,Ha $\rightarrow$ $-0.2256$\,Ha; Table~\ref{tab:Ecorr}).
Again, the inset entropy values increase in tandem with the energy gains, underlining that the rotated basis helps NNCI expose and capture correlation with fewer determinants.\\

\paragraph{\ce{H2O} and \ce{CO} -}
For \ce{H2O} and \ce{CO} we report a single NO set (NO@B$_1$). In both cases, the $E_\mathrm{corr}(N_{\rm orb},N_{\rm det})$ landscape shifts left relative to Hartree-Fock (Figs.~\ref{fig:E_corr_H2O_molecular_orbitals} and \ref{fig:E_corr_CO_molecular_orbitals}).
The best energies improve by $\sim\!7.7$\,mHa for \ce{H2O} ($-0.2193$\,Ha $\rightarrow$ $-0.2270$\,Ha) and by $\sim\!15.7$\,mHa for \ce{CO} ($-0.2965$\,Ha $\rightarrow$ $-0.3122$\,Ha), see Table~\ref{tab:Ecorr}.
We note that \ce{CO} also has the largest \emph{absolute} $|E_\mathrm{corr}|$ among the studied molecules, indicating it is least well described at the mean-field level.
Consistently, for \ce{CO} we find the biggest improvement from aligning the single-particle basis with the correlated occupations.
For both \ce{H2O} and \ce{CO} the off-diagonal weight $W_\mathrm{off}$ follows the same qualitative trend as for the previous cases.

\begin{table}[h!]
\begin{tabular}{lcccc}
\toprule
Molecule & \ce{C3H8} & \ce{NH3} & \ce{H2O} & \ce{CO} \\
\midrule
min($E_\mathrm{corr}^\mathrm{MO}$)  & -0.2895 & -0.2217 & -0.2193 & -0.2965 \\
min($E_\mathrm{corr}^\mathrm{bNO}$) & -0.2913 & -0.2243 & --      & --      \\
min($E_\mathrm{corr}^\mathrm{NO}$)  & -0.2954 & -0.2256 & -0.2270 & -0.3122 \\
$E_\mathrm{corr}$\textsuperscript{a} & -0.2140 & -0.2080 & -0.2160 & -0.2150 \\
\bottomrule
\end{tabular}
\caption{Correlation energy $E_\text{corr}$ in Hartree for different molecules. Reference values\textsuperscript{a} are given for comparison.}
\label{tab:Ecorr}
\begin{flushleft}
\footnotesize\textsuperscript{a}\,These reference calculations \cite{Gao2024} do not use the frozen-core approximation.
\end{flushleft}
\end{table}

Across all molecules, the NO transformation delivers a robust improvement of the $E_\mathrm{corr}$ landscape indicating an efficiency increase w.r.t. the number of required Slater determinants: (i) the variational expansion reaches a given $E_\mathrm{corr}$ with fewer determinants at fixed $N_{\rm orb}$, and (ii) smaller orbital spaces suffice for a given determinant budget.
Moreover, we observe that a \emph{one‑shot} NO (NO@B$_1$) is often a good default that captures most of the efficiency gain (e.g., \ce{NH3}), while a \emph{two‑shot} update (NO@B$_2$) can be worthwhile for larger systems where the one-shot update does not provide the required accuracy.

\section{Conclusions}
In conclusion, we have assessed the effect of using approximate NOs - obtained from intermediate many-body solutions - as the single-particle basis for neural-network-assisted selective configuration interaction. 
Across four benchmark molecules (\ce{H2O}, \ce{NH3}, \ce{CO}, \ce{C3H8}), NOs consistently lower the computational effort needed to reach a given correlation-energy accuracy and improve the best energies attained within a fixed scan of $(N_{\rm orb},N_{\rm det})$. 
In all systems, the $E_{\rm corr}(N_{\rm orb},N_{\rm det})$ landscapes w.r.t. canonical Hartree-Fock orbitals improve significantly, indicating a more favorable determinant-orbital trade-off in the NO basis.

Practically, we found that a single comparatively cheap NO update is a robust default that captures most of the benefit - a second update (\mbox{NO@B$_2$}) is worthwhile for larger systems or higher accuracy goals, as the total error scales with system size and maintaining chemically meaningful absolute energy differences thus requires higher relative accuracy. 
These findings provide a minimal, drop-in recipe to reduce determinant budgets while improving correlation energies in ML-accelerated SCI workflows.

\begin{acknowledgements}
This work was supported by the Icelandic Research Fund (grant nos.\ 2511544 and 2410644) and the University of Iceland Research Fund. G.L. acknowledges support from the ERC under the European Union's Horizon Europe research and innovation programme (grant no. 101166044, project NEXUS). 
Views and opinions expressed are however those of the author(s) only and do not necessarily reflect those of the European Union or ERC Executive Agency. 
Neither the European Union nor the granting authority can be held responsible for them. Computer resources, data storage, and user support were provided by the Icelandic Research e-Infrastructure (IREI), funded by the Icelandic Infrastructure Fund. 
The authors further gratefully acknowledge the scientific support and HPC resources provided by the Erlangen National High Performance Computing Center (NHR@FAU) of the Friedrich-Alexander-Universität Erlangen-Nürnberg (FAU). 
P.B.\ gratefully acknowledges the ARTEMIS funding via the QuantERA program of the European Union provided by German Federal Ministry of Education and Research under the grant 13N16360 within the program ``From basic research to market''. 
Y.L.A.S.\ and P.B.\ acknowledge support by the Max Planck Society.
\end{acknowledgements}

\newpage

\bibliography{main_references}

\end{document}